\documentclass[12pt,preprint,showpacs,preprintnumbers,amsmath,amssymb]{revtex4}

\usepackage{graphicx}
\usepackage{dcolumn}
\usepackage{bm}

\begin{document}

\title{Inelastic Scattering and Interactions of Three-Wave Parametric Solitons}


\author{ Matteo Conforti, Fabio Baronio}

\affiliation{ Dipartimento di Elettronica per l'Automazione,
Universit\`a di Brescia, \\ Via Branze 38, 25123 Brescia, Italy
}%

\author{Antonio Degasperis}
\affiliation{Dipartimento di Fisica, Istituto Nazionale di Fisica Nucleare,
Universit\`a ``La Sapienza'', P.le A. Moro 2, 00185 Roma, Italy}%

\author{Stefan Wabnitz}

\affiliation{ Laboratoire de Physique, Universit\'e de Bourgogne,
UMR CNRS 5027, 9 Av. A. Savary, 21078 Dijon, France
}%

\date{\today}

\begin{abstract}
We study the excitation, decay and interactions of novel, velocity
locked three-wave parametric solitons in a medium with quadratic
nonlinearity and dispersion. We analytically describe the
particle-like scattering between stable or unstable soliton
triplets with linear waves in terms of explicit solutions
featuring accelerated or decelerated solitons.
\end{abstract}

\maketitle

Three-wave resonant interactions (TWRI) are widespread in various
branches of physics, as they describe the resonant mixing of waves
in weakly nonlinear and dispersive media. The TWRI model is
typically encountered in the description of any conservative
nonlinear medium where the nonlinear dynamics can be considered as
a perturbation of the linear waves solution, the lowest-order
nonlinearity is quadratic in the field amplitudes and the
phase-matching (or resonance) condition is satisfied.
Solutions to the TWRI have been known for a long time
\cite{armstrong70,zakharov73,nozaki73,nozaki74,nozaki74bis,kaup76,kaup79},
and extensive applications are found in nonlinear optics
(parametric amplification, frequency conversion, stimulated Raman
and Brillouin scattering), plasma physics (laser-plasma
interactions, radio frequency heating, plasma instabilities),
acoustics (light-acoustic interactions), fluid dynamics
(interaction of water waves) and solid state physics (wave-wave
scattering). Soliton solutions of TWRI are of particular interest
in the study of coherent energy transport and frequency
conversion. Indeed, solitons behave as particles: as a result,
different waves belonging to the same soliton may propagate locked
together as a single entity. Such effect has no counterpart for
linear waves
\cite{armstrong70,taranenko92,ibragimov96,ibragimov97,ibragimov98,picozzi98}.
In a related paper \cite{deg06}, we discovered a novel
multi-parameter TWRI soliton family consisting of a triplet of
bright-bright-dark waves (or simulton) that travel with a common
velocity. We could also identify the conditions for stable (fast)
and unstable (slow) solitons. The fate of the unstable solitons
however remained unknown.

In this Letter, the solution of this problem is fully reported by
giving, strikingly enough, an analytic exact solution which
describes the entire evolution of the unstable solitons. In fact,
we reveal and explore a novel consequence of the particle-like
nature of TWRI simultons (TWRIS), namely their inelastic
scattering with particular linear waves. Such phenomenon is
associated with the excitation (decay) of stable (unstable)
simultons by means of the absorption (emission) of the energy
carried by an isolated linear pulse. The decay (excitation) of
simultons is associated with their speed-up (slowing-down) and
creation of another triplet with complementary stability
properties. As shown below, such processes are exactly described
in terms of an analytical higher-order soliton solution with
varying speed, or boomeron. The present TWRIS scattering process
may be pictured as the interaction of radiation with a two-level
atomic system: transitions among excited and ground soliton states
are induced by the absorption and spontaneous emission of a wave.

The coupled partial differential equations that rule TWRI in (1 + 1)
dimensions read as \cite{zakharov73}:
\begin{eqnarray}\label{3wri}
\nonumber E_{1t}-V_1 E_{1z}&=&\phantom{-} E_2^*E_3^*,\\ E_{2t}-V_2
E_{2z}&=&-E_1^*\,E_3^*,\\ \nonumber E_{3t}-V_3
E_{3z}&=&\phantom{-} E_1^*\,E_2^*,
\end{eqnarray}
where the subscripts $t$ and $z$ denote derivatives in  the
longitudinal and transverse dimensions, $E_n=E_n(z,t)$ are the
complex wave amplitudes with velocities $V_n$, and $n=1,2,3$. We
chose here $V_1>V_2>V_3$ which, together with the above choice of
the signs before the quadratic terms, entails the non-explosive
character of the three-wave interaction \cite{kaup79}. In the
following, with no loss of generality, we shall consider
Eqs.(\ref{3wri}) in a reference frame with $V_3=0$. Since we
consider resonant interactions, the frequencies and momenta of the
three waves must satisfy the prescriptions
$\omega_1+\omega_2+\omega_3=0$ and $k_1+k_2+k_3=0$.
%

The TWRI equations (\ref{3wri}) represent an infinite-dimensional
Hamiltonian system, which conserves the Hamiltonian, the sum of
the energies of waves $E_1$ and $E_2$, the sum of the energies of
waves $E_2$ and $E_3$, and the total transverse momentum (see Ref.
\cite{deg06} for details).

\begin{figure}
 \begin{center}
          \includegraphics[width=8cm]{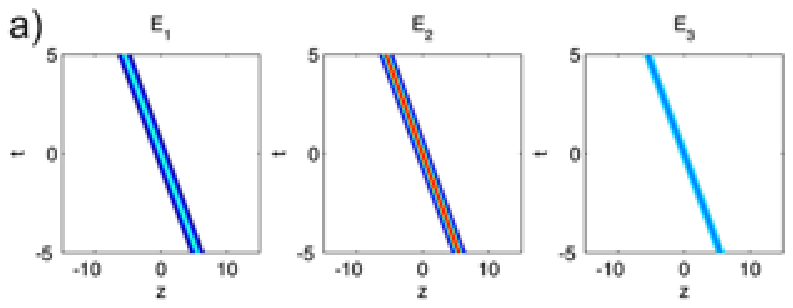}\\
          \includegraphics[width=8cm]{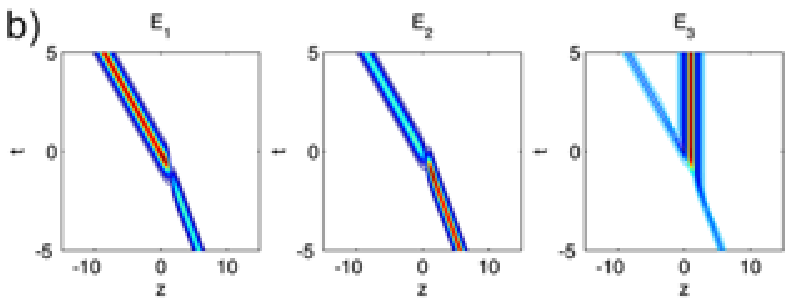}
 \end{center}
    \caption{a) Analytical solution; b) numerical propagation of an unstable TWRIS,
which coincides with a boomeron solution.
    Here $V_1=2,V_2=1$. The simulton velocity is $V=1.1$ ($V<V_{cr}\approx 1.3$).}\label{runs}
\end{figure}

Equations (\ref{3wri}) exhibit a three-parameter family of
simulton solutions in the form of bright-bright-dark triplets that
travel with a common, or locked velocity $V$ \cite{deg06}. The
most remarkable physical property of these simultons is that their
speed $V$ may be continuously varied by means of adjusting the
energy of the two bright pulses. The propagation stability
analysis of TWRIS reveals that a triplet is no longer stable
whenever its velocity $V$ decreases below a well defined
(critical) value, namely $V<V_{cr}=2V_1V_2/(V_1+V_2)$
\cite{deg06}. As an example, in Fig. \ref{runs}(a) we show the
contour plot of the amplitude of the three waves that compose an
unstable simulton. These plots should be compared with Fig.
\ref{runs}(b), obtained from the numerical propagation with an
initial (i.e., at $t=-5$) condition given by the exact simulton
solution of Fig. \ref{runs}(a). The results of Fig. \ref{runs}
illustrate that an unstable simulton with $V<V_{cr}$ decays into a
stable simulton with $V>V_{cr}$. This process is accompanied by
the emission of an isolated pulse in wave $E_3$. It is quite
remarkable that the simulton decay and wave emission as it is
numerically observed in Fig. \ref{runs}(b) may be exactly
reproduced in terms of an analytical higher-order soliton solution
with varying speed, or boomeron. Such solution was found by means
of the techniques described in Ref.\cite{calogero05}, and it can
be expressed as
\begin{subequations}\label{boomeron}
\begin{equation}\label{E1}
E_1=\frac{2p V_2}{\Delta} \sqrt{\frac{2V_1}{V_1-V_2}}e^{iq_1
z_1}(H_+^* e^{-i\theta}-H_-^* e^{i\theta}),
\end{equation}
\begin{equation} \label{E2}
\begin{split}
E_2=\frac{2p V_1}{\Delta} \sqrt{\frac{2V_2}{V_1-V_2}}e^{iq_2
z_2}\Big (\sqrt{{(1-Q)}{/(1+Q)}}\\ H_+ e^{i(\beta + \theta)}
-\sqrt{{(1+Q)}/{(1-Q)}} H_- e^{-i(\beta + \theta)} \Big),
\end{split}
\end{equation}
\begin{equation} \label{E3}
\begin{split}
E_3=a\sqrt{V_1 V_2} e^{iq_3 z_3}-\frac{\Delta}{4p}
\big(\frac{V_1-V_2}{V_1 V_2} \big ) E_1^* E_2^*,
\end{split}
\end{equation}
\end{subequations}
where
\begin{align*}
   &\Delta= 1+\frac{|H_+|^2}{1+Q}+ \frac{|H_-|^2}{1-Q}-2 \cos(\beta) \mathcal{R}e (H_+ H_-^*
   e^{i(\beta+2\theta)})\,\,, \\
   &H_{\pm} (z,t)= e^{(-B_{\pm} +i \chi_{\pm} )z} \, e^{\frac{-2V_1V_2}{V_1-V_2}(p-ik)t}\,,  \\
   & \omega = -2k \frac{V_1 V_2}{V_1-V_2}, \ \ \
   \chi _{\pm}= k\big (\frac{V_1+V_2}{V_1-V_2} \mp \frac{1}{Q}\big
   ), \\
   &B_{\pm}= p\big (\frac{V_1+V_2}{V_1-V_2} \mp Q \big), \ \ \ \tan(\beta)=k/(pQ), \\
   & Q= \frac{1}{p}\sqrt{ \frac12 [\,\,r + \sqrt{r^2+4k^2p^2}\,\,]},
   \,r=p^2-k^2-a^2\,\,, \\
   & q_n=q(V_{n+1}-V_{n+2}), \\
   & z_n =z+V_n t, \, n=1,2,3 \,mod(3) .
 \end{align*}

It is worth noting that the above solution depends upon seven real
parameters $V_1,V_2,p,k,q,a,\theta$. From the definition of $Q$,
it is apparent that these parameters must be chosen in such a way
that if $k=0$, then $p^2>a^2$.

The analytical solution (\ref{boomeron}), while rather complicate
at intermediate times, asymptotically consists of one or two
coherent structures.
 In fact, let us consider first the decay process: if we assume $p<0$,
 for negative large $t$
($t\rightarrow -\infty$) the boomeron is asymptotically composed
of two bright pulses ($E_1$, $E_2$) and a kink-like pulse ($E_3$)
travelling with the locked velocity $V_{i}$. If instead $t$ is
large and positive ($t\rightarrow +\infty$) the boomeron is
composed of two bright pulses ($E_1$, $E_2$) and a kink-like pulse
($E_3$) traveling at the locked velocity $V_{f}$ ($V_{f}>V_{i}$),
plus another pulse ($E_3$) that travels with the linear group
velocity $V_3$. The velocities $V_{f}$ and $V_{i}$ can be
calculated from (\ref{boomeron}):
\begin{align}
    V_{i} =\frac{ 2 V_1 V_2}{
V_1+V_2 - Q(V_1-V_2)},\\
    V_{f} =\frac{ 2 V_1 V_2}{
V_1+V_2 + Q(V_1-V_2)}.
 \end{align}
The triplet travelling at very large $|t|$ with the locked
velocity $V_{i}$ ($V_{f}$)  is itself an exact solution of
Eqs.(\ref{3wri}), namely it is the unstable (stable) TWRIS as
presented in Ref.\cite{deg06}. Therefore the boomeron solution
(\ref{boomeron}) provides the exact description of the decay from
unstable into stable solitons.
%
%

Let us consider next the situation where a stable TWRIS collides
with an isolated pulse in the wave $E_3$, namely the excitation by
absorption. Once again, this scattering process is exactly
described by the boomeron solution (\ref{boomeron}), and it leads
to the excitation of an unstable TWRIS, induced by the absorption
of the isolated wave $E_3$. Indeed, whenever $p>0$ and $t$ is very
large and negative, the boomeron (\ref{boomeron}) is composed of a
triplet consisting of two bright pulses (in waves $E_1$, $E_2$)
and a kink-like pulse (in wave $E_3$), all traveling with the same
velocity $V_{i}$, plus an isolated pulse in wave ($E_3$) that
travels with the linear group velocity $V_3$. The triplet and the
isolated pulse collide and, as a result, the pulse in $E_3$ is
completely absorbed by the triplet. Finally, for very large and
positive $t$ the boomeron consists of a single triplet formed by
two bright pulses (in waves $E_1$, $E_2$) and a kink-like pulse
(in wave $E_3$), again traveling together with the velocity
$V_{f}$ ($V_{f}< V_{i}$). Note that the asymptotic boomeron
triplets traveling with velocities $V_{i}$ and $V_{f}$ can be
analytically mapped into the stable and unstable TWRIS as given in
\cite{deg06}. In conclusion, the analytical solution
(\ref{boomeron}) with $p>0$ provides the exact description of the
excitation of an unstable TWRIS as a result of the inelastic
collision between a stable TWRIS and a linear wave packet.

\begin{figure}
 \begin{center}
          \includegraphics[width=8cm]{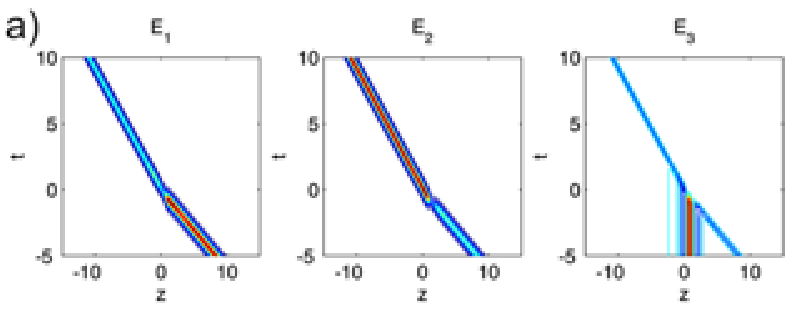}\\
          \includegraphics[width=8cm]{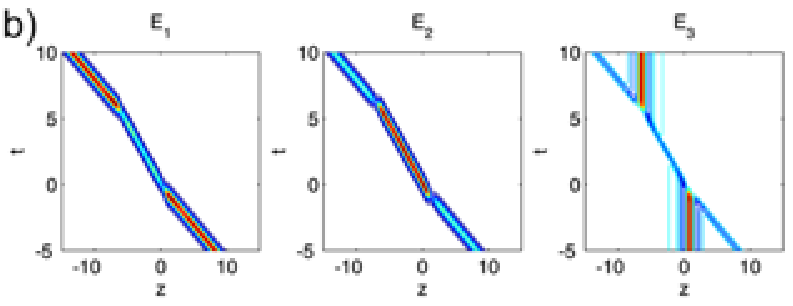}
 \end{center}
    \caption{a) Analytical boomeron solution describing the collision of a stable TWRIS with a single pulse in
    wave $E_3$. Parameters are $V_1=2, V_2=1, V_3=0, p=1, a=1, k=0.5, q=1,
    \theta=\pi/6$. The triplet velocities are $V_{i}=1.8$ and $V_{f}=1.1$ ($V_{cr}\approx 1.3$).
    b) Numerical double scattering process.}\label{runs2}
\end{figure}

Figure \ref{runs2}(a) displays the analytical boomeron solution
corresponding to the collision between a stable TWRIS and a pulse
in wave $E_3$. Whereas Fig. \ref{runs2}(b) shows the inelastic
scattering of the TWRIS and the linear wave as numerically
computed by integrating the equations (\ref{3wri}) with the
initial data at $t=-0.5$ equal to the solution of Fig.
\ref{runs2}(a). As it can be seen in Fig. \ref{runs2}(b), the
excited unstable TWRIS has a finite lifetime since it eventually
decays into a stable or ground state TWRIS via the emission of
another linear wave. It is worth noting that both the excitation
and the decay processes may described by properly adjusting the
parameters of Eqs.(\ref{boomeron}).
%
%

%
%

The dynamics of the scattering between TWRIS and linear waves is
analogous to the interaction between radiation and a two-level
atom. Indeed, transitions between excited and ground soliton
states are induced by the absorption and spontaneous emission of a
linear pulse in the wave $E_3$.


%
Let us now briefly discuss the role of the various parameters in
Eqs. (\ref{boomeron}). Two of these parameters (i.e. the
velocities $V_1$ and $V_2$) are fixed by the linear dispersive
properties of the medium. We are thus left with five independent
real parameters, namely $p,k,q,a, \theta$ (with the restrictions
$a>0$ and $0\le\theta<2\pi$). 
We point out that our discussion above implies that the
specification of these parameters allows one to define the
properties of both unstable and stable TWRIS since these solitons
result as asymptotic states of the analytic boomeron expression
(\ref{boomeron}) in the limit as $|t|\rightarrow\infty$. The
parameter $p$ is associated with the rescaling of the wave
amplitudes, and of the coordinates $z$ and $t$. Whereas $a$
measures the amplitude of the kink background in wave $E_3$. The
value of $k$ is related to the soliton wave—number. The parameter
$q$ simply adds a phase shift which is linear in both $z$ and $t$.
Finally, $\theta$ fixes the shape of the stationary kink pulse
$E_3$. By adjusting the various degrees of freedom of the boomeron
family of solutions (\ref{boomeron}), one may foresee the
dynamical reshaping of the amplitude, phase, and velocity of the
TWRIS, as well as fully describe the process of energy exchange
among the three waves.

\begin{figure}
 \begin{center}
          \includegraphics[width=8cm]{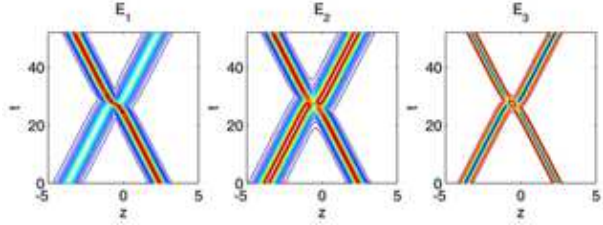}
 \end{center}
    \caption{Collision of two stable TWRIS with different
    velocities. Fast simulton $V=1.9$, slow simulton $V=1.7$.
    Simulation is performed in reference frame moving at velocity
    $V_{ref}=1.8$.
    }\label{collis}
\end{figure}
\begin{figure}
 \begin{center}
          \includegraphics[width=8cm]{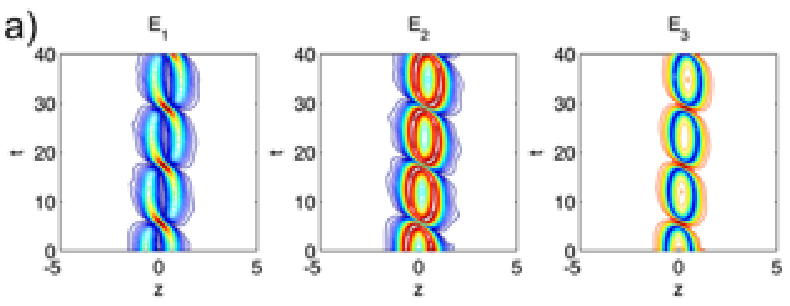}\\
          \includegraphics[width=8cm]{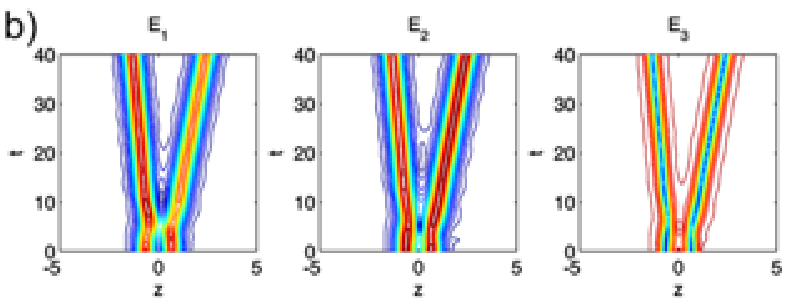}
 \end{center}
    \caption{a) Collision of two equal and in-phase stable TWRIS with the same
    velocity $V=1.8$; b) collision of two equal and $\pi/4$ out-of-phase stable TWRIS with the same velocity.
    Simulations are performed in reference frame moving at velocity $V_{ref}=1.8$.}\label{collis2}
\end{figure}
In order to emphasize the novel and striking features of the
scattering between TWRIS and linear waves, let us briefly consider
now, the collisions between different TWRIS. Since
Eqs.(\ref{3wri}) are completely integrable, interactions between
two initially well-separated TWRIS do not modify the shapes of
triplets that emerge after the collision. Indeed, the numerical
simulation of Fig. \ref{collis} shows that two TWRIS with
different velocities penetrate and cross each other with no change
of their shapes. The only effect of the interaction is a spatial
shift and a phase shift, as it happens with ordinary bright TWRI
solitons \cite{ibragimov97}. However, in a manner similar to cubic
nonlinear Schr\"odinger solitons (\cite{gord83,koda87},
\cite{hase95} and references therein), whenever the initial
simulton separation is reduced, complex interaction phenomena may
take place owing to the excitation of higher order soliton
solutions. For example, Fig. \ref{collis2}(a) shows that two equal
and in-phase TWRIS with the same velocity attract each other and
periodically collapse. Whereas Fig. \ref{collis2}(b) shows that a
repulsive force exists between two equal and out-of-phase solitons
with the same velocity (the phase difference $\alpha$ between the
two solitons is imposed by multiplying the wave $E_1$,
respectively $E_2$, of one of the solitons by the phase factor
$\exp[{i\alpha}]$, respectively $\exp[{-i\alpha}]$). In this case,
two distinct TWRIS moving with different velocities emerge from
the initial collision. Hence TWRI solitons may cross, attract or
repel each other depending on their initial separation, velocity
difference, and relative phase.

In conclusion, we described in terms of analytical solutions the
scattering process of three-wave simultons and linear waves. An
unstable simulton decays into a stable simulton by accelerating
its speed and emitting an isolated pulse. Moreover, a stable
triplet may be excited into an unstable simulton by slowing down
as a result of the absorption of a linear wave. Finally, simultons
with different speeds are stable upon collision, and simultons
with equal speeds interact with each other in a way which is
strongly dependent upon their initial relative phase.

\

\end{document}